\documentclass[journal=jpclcd,manuscript=letter]{achemso}

\usepackage[version=3]{mhchem} 
\usepackage{graphicx}
\usepackage[sort&compress,numbers,super]{natbib}

\author{Ruo Xi Yang}
\affiliation{Department of Chemistry, University of Bath, Claverton Down, BA2 7AY, UK}
\alsoaffiliation{Department of Materials, Imperial College London, Exhibition Road, London SW7 2AZ, UK}

\author{Jonathan M. Skelton}
\affiliation{Department of Chemistry, University of Bath, Claverton Down, BA2 7AY, UK}

\author{E. Lora da Silva}
\affiliation{Department of Chemistry, University of Bath, Claverton Down, BA2 7AY, UK}

\author{Jarvist M. Frost}
\affiliation{Department of Chemistry, University of Bath, Claverton Down, BA2 7AY, UK}
\alsoaffiliation{Department of Materials, Imperial College London, Exhibition Road, London SW7 2AZ, UK}

\author{Aron Walsh}
\email{a.walsh@imperial.ac.uk}
\affiliation{Department of Materials, Imperial College London, Exhibition Road, London SW7 2AZ, UK}
\alsoaffiliation{Department of Materials Science and Engineering, Yonsei University, Seoul 03722, Korea}

\title[An \textsf{achemso} demo]{ Spontaneous Octahedral Tilting in the Cubic Inorganic Caesium Halide Perovskites \ce{CsSnX3} and \ce{CsPbX3} (X = \ce{F}, \ce{Cl}, \ce{Br}, \ce{I})}

\begin{tocentry}
\scalebox{1}{\includegraphics{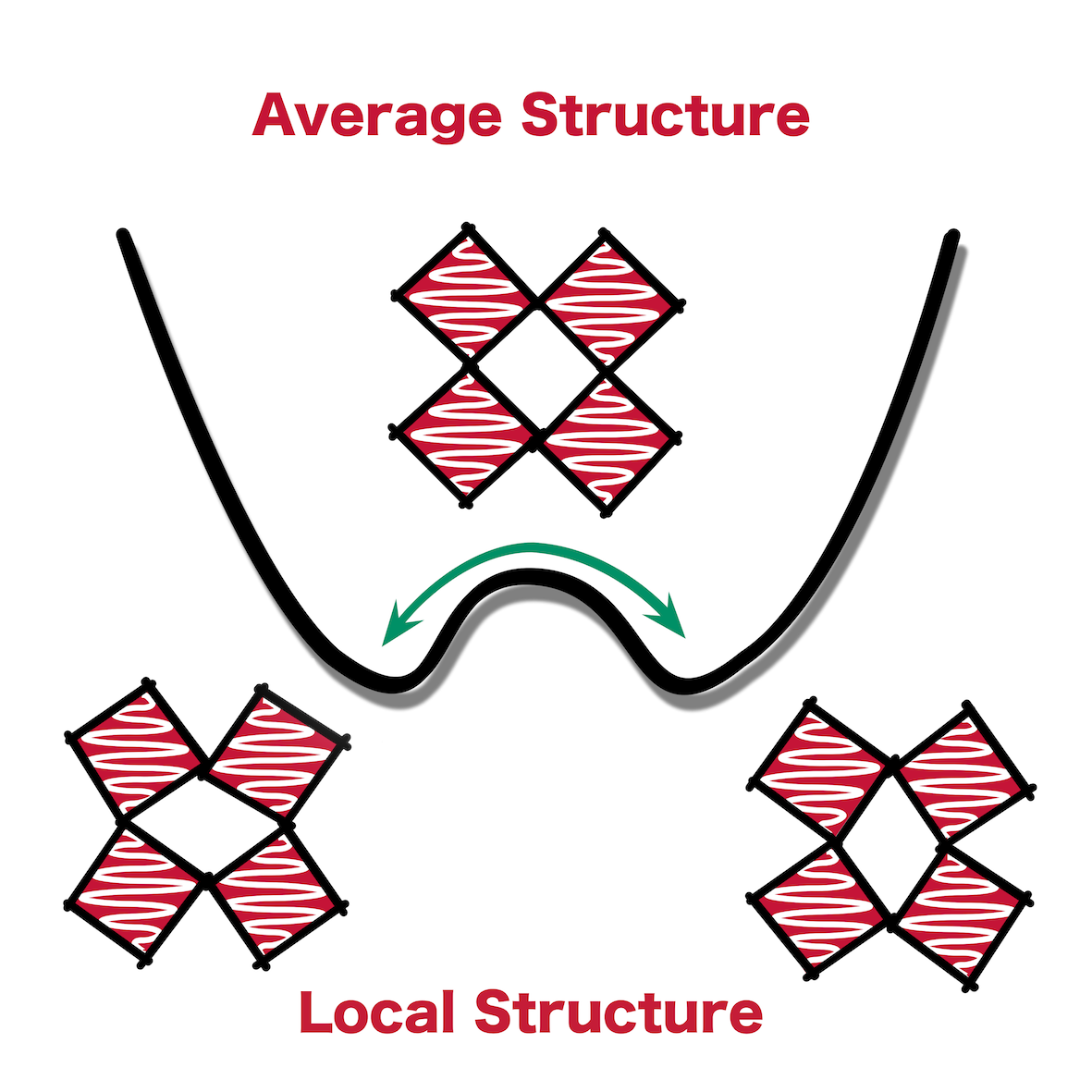}}
\end{tocentry}

\begin{document}


\begin{abstract}
The local crystal structures of many perovskite-structured materials deviate from the average space group symmetry.  We demonstrate, from lattice-dynamics calculations based on quantum chemical force constants, that all the caesium-lead and caesium-tin halide perovskites exhibit vibrational instabilities associated with octahedral titling in their high-temperature cubic phase. Anharmonic double-well potentials are found for zone-boundary phonon modes in all compounds with barriers ranging from 108 to 512 meV. The well depth is correlated with the tolerance factor and the chemistry of the composition, but is not proportional to the imaginary harmonic phonon frequency. We provide quantitative insights into the thermodynamic driving forces and distinguish between dynamic and static disorder based on the potential-energy landscape.  A positive band gap deformation (spectral blueshift) accompanies the structural distortion, with implications for understanding the performance of these materials in applications areas including solar cells and light-emitting diodes.
\end{abstract}


Since the discovery of photoconductivity in the caesium-lead halides (\ce{CsPbX3}; \ce{X} = \ce{Cl}, \ce{Br}, \ce{I}),\cite{Moller1958} the semiconducting properties of halide perovskites have attracted significant research attention, including analogous compounds based on tin.\cite{Huang2013c,Chung2012b}
Interest has since expanded to the hybrid organic-inorganic perovskites, with applications ranging from field-effect transistors,\cite{Mitzi1994} photovoltaics,\cite{Kojima2009, Lee2012} and light-emitting diodes.\cite{Ning2015a}
This family of materials display a unique combination of physical and chemical properties, including fast ion and electron transport, long minority-carrier diffusion lengths, and high quantum efficiencies.

The crystallography of halide perovskites dates back to the 1950s, where the high-temperature crystal structures of the \ce{CsPbX3} series were determined to be the prototypical cubic-perovskite structure (space group $Pm\bar{3}m$).\cite{Moller1958}
The crystal structure consists of Cs in a cuboctahedral cavity at the centre of a corner-sharing lead halide octahedral network. 
The same high-temperature structure was also reported for the organic-inorganic \ce{CH3NH3PbX3} series.\cite{Weber1978}
In all cases, phase transitions to lower-symmetry perovskite phases are observed at lower temperatures, e.g. in \ce{CsPbCl3} there is a transition to a tetragonal phase at 320 K, an othorhombic phase at 316 K, and a monoclinic phase at 310 K.\cite{Fujii1974}

In the 1970s, Poulsen \textit{et al.} determined the room-temperature structure of \ce{CsSnCl3} to be monoclinic ($P2_1/n$ type), and identified a phase transition to a higher-symmetry structure at 393 K.\cite{Poulsen1970}
An X-ray diffraction (XRD) study of \ce{CsSnBr3} determined the structure to be cubic at room temperature, but symmetry lowering was observed as the temperature was reduced.\cite{Barrett1971a}
More recently, temperature-dependent synchrotron XRD experiments determined \ce{CsSnI3} to be cubic at 500 K, with tetragonal and orthorhombic phases observed at lower temperatures.\cite{Chung2012b}
It was suggested that the phase transitions are associated with the 5$s^2$ lone electron pair of \ce{Sn}, and the consequential distortion of the corner-sharing octahedral \ce{BX3} framework.

Despite numerous crystallographic studies on lead- and tin-based perovskites, the nature of the high-temperature cubic phases of the these compounds has received less attention.
Analysis of the X-ray pair distribution functions of \ce{CH3NH3SnBr3} suggested that the local cubic symmetry was broken, with significant distortions of the octahedral network.\cite{Worhatch2008}
It was recently confirmed from both inelastic X-ray scattering and neutron total scattering that the cubic phase of \ce{CH3NH3PbI3} is also symmetry broken.\cite{Beecher2016,Druzbicki2016}
These observations have been associated with the rotational disorder of the molecular \ce{CH3NH3+} cation.
For inorganic halide perovskites, this molecular disorder is absent, and instead the disorder in the cubic inorganic halide perovskites should be due solely to the flexibility associated with the inorganic octahedral network.

It is also interesting to note that many quantum dots and nanoparticles of halide perovskites have been reported to adopt a cubic structure at room temperature.\cite{Song2015b,Protesescu2015}
It was unclear initially whether the stability of the cubic phase was due to surface effects, lattice strain, or phonon confinement. However, a recent X-ray total scattering study of colloidal \ce{CsPbX3} (X = \ce{Cl}, \ce{Br}, \ce{I}) nanocrystals provided the first evidence that the local structure is not cubic, but consists of domains with orthorhombic tilting.\cite{Bertolotti2017}

In this Letter, we demonstrate that spontaneous octahedral tilting is common to caesium-lead and caesium-tin halide perovskites.
Through first-principles lattice-dynamics calculations, we assess the chemical and thermodynamic driving forces for these instabilities.
Double well potentials are found for ``soft'' phonon modes in all cases, with barrier heights ranging from 108 to 512 meV.
We also show that octahedral tilting results in a positive band gap deformation, indicating that local symmetry breaking would lead to a larger band gap than anticipated from the regular cubic perovskite structure.

\textit{Octahedral tilting in perovskites:}
The aristotype cubic \ce{ABX3} perovskite structure is usually only observed at high temperature, while at lower temperatures a group of lower-symmetry phases, including tetragonal, orthorhombic, monoclinic, and rhombohedral, are found.
With reference to the cubic phase, the associated phase transitions are driven by a range of symmetry-breaking lattice distortions.
The phase diversity of perovskites can be qualitatively explained using the concept of the tolerance factor introduced by Goldschmidt \cite{Goldschmidt1926}, where
\begin{equation}
\alpha = \frac{r_A + r_X}{\sqrt{2}(r_B + r_X)}
\end{equation}
with $r_A$, $r_B$ and $r_X$ being the ionic radii for the A, B and X atom, respectively. Values of $\alpha < 1$ are usually associated with octahedral titling, due to the A cation being smaller than is optimal for bonding with the \ce{BX3} framework.
This is the case for the majority of compounds considered here, which explains the experimentally observed \textit{Pnma} ground-state structures of \ce{CsPbCl3} and \ce{CsSnI3}.\cite{Ahtee:a18798,Chung2012b}

Glazer developed a simple classification system to describe the octahedral tilting in perovskites and to relate it to phase transitions.\cite{Glazer1972}
In the Glazer notation, octahedral tilting is described as a linear combination of in-phase and out-of-phase rotations along the crystallographic axes: for example, the notation $a^0b^-c^-$ indicate two out-of-phase tilts along the [010] and [001] directions ($b$ and $c$ axes) with distinct tilt angles.
More recently, Stokes \textit{et al.} provided a group-theoretical description of the relationships between different tilt systems, and also considered \ce{B} cation displacements within the octahedra,\cite{Howard1998,Howard2006} where the distortion is expressed with irreducible representations such as $M_3^+$ or $R_4^+$.
For example, the cubic ($Pm\bar{3}m$) to tetragonal ($P4/mbm$) phase transition of \ce{CsPbCl3} can be described as $a^0a^0a^0 \rightarrow a^0a^0c^+$ tilting or, equivalently, as the condensation of an $M_3^+$ phonon mode.\cite{Fujii1974} 
Woodward provided insights into the stabilizing chemical forces based on the bonding environment and crystal structure (tolerance factor) of specific oxide compounds,\cite{Woodward1997,Woodward1997a}
which have recently been applied to understanding the octahedral titling preferences of iodide and bromide perovskites.\cite{young2016octahedral}

A search of the Inorganic Crystal Structure Database (ICSD),\cite{Bergerhoff1983} summarised in Table \ref{t:phase_table}, reveals that \ce{CsPbCl3}, \ce{CsPbBr3}, \ce{CsPbI3}, \ce{CsSnI3} adopt cubic phases (space group $Pm\bar{3}m$) above room temperature\cite{Ahtee:a18798,sakata1979,Trots2008,Chung2012b}, while \ce{CsSnBr3} and \ce{CsSnCl3} has been reported to be cubic at room temperature (from XRD) and \ce{CsPbF3} at 187 K (from neutron diffraction).\cite{Barrett1971a,Donaldson1975a,Berastegui2001} 
Recently, low-frequency Raman spectroscopy has shown that the cubic phase of \ce{CsPbBr3} determined by XRD fluctuates at short timescale between different lower symmetry phases but appear to be cubic on average, due to the structural flexibility, a phenomenon that could be present in other halide perovskites. \cite{Yaffe2017} 

\begin{table}[]
\centering
\begin{tabular}{ccccc}
\hline
			&	Cubic ($Pm\bar{3}m$)	&	Tetragonal ($P4/mbm$)	&	Orthorhombic ($Pnma$)	&	Monoclinic($P2_1/n$) \\ \hline
\ce{CsSnF3} &  & & & \\
\ce{CsSnCl3} & 293 K \cite{Barrett1971a} & & & $<$ 293 K \cite{Barrett1971a}	\\ 
\ce{CsSnBr3} & 292 K \cite{Donaldson1975a},300 K\cite{Fabini2016b} & 270 K \cite{Fabini2016b}& 100 K \cite{Fabini2016b} & $<$ 292 K \cite{Donaldson1975a}	\\ 
\ce{CsSnI3}  & 500 K \cite{Chung2012b}, 446 K\cite{Yamada1991a} & 380 K \cite{Chung2012b}, 373 K\cite{Yamada1991a}& 300 K \cite{Chung2012b,Yamada1991a} & \\ 
\ce{CsPbF3}	 & 186 K \cite{Berastegui2001} & & & \\ 
\ce{CsPbCl3} & 320 K \cite{Fujii1974} & 315 K \cite{Fujii1974} & 310 K \cite{Fujii1974} & $<$ 310 K \cite{Fujii1974}\\ 
\ce{CsPbBr3} & 403 K \cite{Hirotsu1974} & 361 K \cite{Hirotsu1974} & $<$ 361 K \cite{Hirotsu1974} & \\ 
\ce{CsPbI3} & 634 K \cite{Trots2008} & & 298 K \cite{Trots2008} \\ \hline
\end{tabular}
\caption{Comparison of known inorganic halide perovskite phases and the temperature above which the phase are observed for each composition.}
\label{t:phase_table}
\end{table}

\textit{Harmonic lattice dynamics:}
We start by computing the harmonic phonon frequencies and dispersions for eight inorganic halide compounds \ce{ABX3} (\ce{A} = \ce{Cs}; \ce{B} = \ce{Sn}, \ce{Pb}; \ce{X} = \ce{F}, \ce{Cl}, \ce{Br}, \ce{I}) in the cubic perovskite structure.
Lattice dynamic calculations were performed using the open-source \textsc{Phonopy}\cite{Togo2015a} package with forces calculated within the Kohn-Sham density-functional theory (DFT) formalism, as implemented in the VASP code\cite{Kresse1996a,Kresse1996}.
Particular attention was given to the convergence of the energy and forces, $1\times10^{-8}$ eV and $1\times10^{-3}$ eV/\AA, respectively; production calculations were performed with the exchange-correlation functional PBEsol,\cite{Perdew2008,Perdew2009} using an $8\times8\times8$ sampling of the electronic Brillouin zone and a plane wave cut-off of 800 eV.
Projector augmented wave\cite{Blochl1994} core potentials (reciprocal space projection) were used with valence $4d^{10}5s^25p^2$ electrons on Sn,  $6s^26p^2$ electrons on Pb, $5s^{2}5p^{6}6s^{1}$ electrons on Cs, and outmost $ns^{2}np^{5}$ electrons on halogen atoms (X).  
Input structures were built in the cubic perovskite structure $Pm\bar{3}m$, and fully relaxed with fixed symmetry.
The phonon frequencies and eigenvectors were determined by finite-displacement calculations with a step size of 0.01 \AA, performed in $2\times2\times2$ supercell expansions of the cubic unit cell.
An imaginary phonon frequency, demonstrated as negative frequency in the phonon band structure, indicates the presence of a structural instability, i.e. the phase is not a true local minimum on the potential energy surface. 
The structure can distort along the pathway determined by the phonon eigenvector to lower the internal energy.  The energy landscape as a function of distortion amplitude is obtained by the code ModeMap.\cite{Skelton2016,Skelton_ModeMap}

\begin{figure}
	\includegraphics[width=\linewidth]{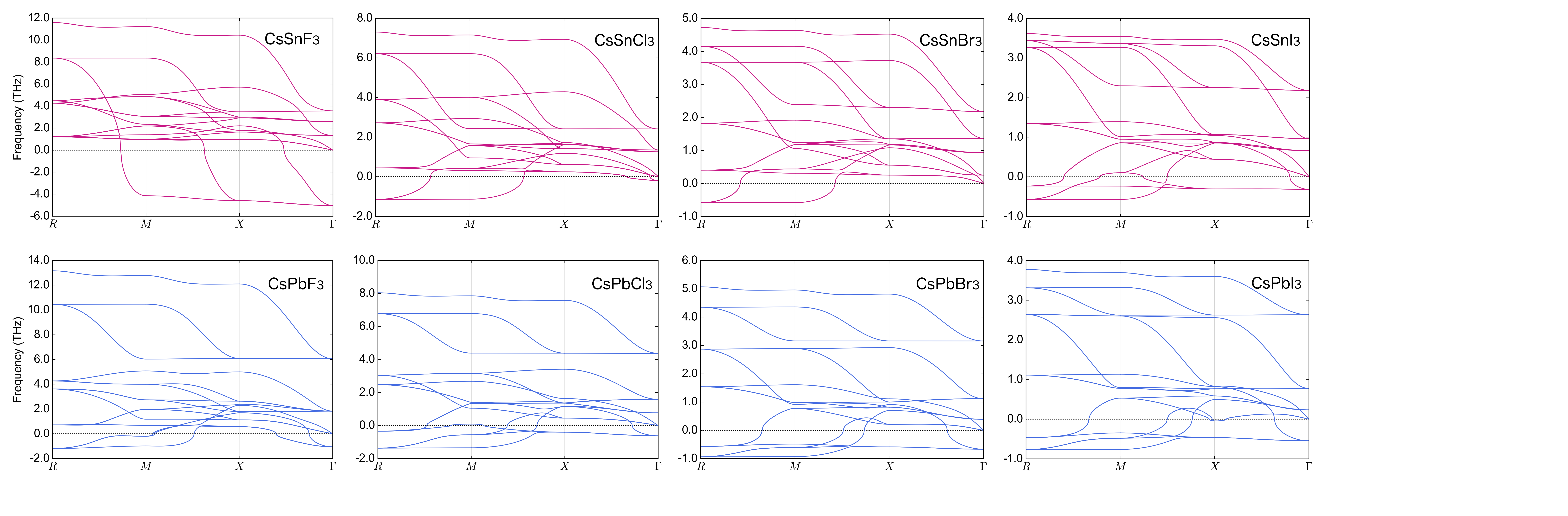}
	\caption{
    Harmonic phonon dispersion of \ce{ABX3} compound in the cubic perovskite structure. The labels correspond to special points in the vibrational Brillouin zone: $\Gamma (0,0,0)$, $X (\frac{1}{2},0,0)$, $M (\frac{1}{2},\frac{1}{2},0)$ and $R (\frac{1}{2},\frac{1}{2},\frac{1}{2})$. Imaginary frequencies are represented by negative numbers on the $y$ axis for convenience of plotting. 
    }
	\label{phonon_band}
\end{figure}

The phonon dispersion of all eight compositions display imaginary frequencies (lattice instabilities) in the phonon Brillouin zones (Fig.\ref{phonon_band}).
Instabilities associated with tilting of the octahedra can be found at the Brillouin zone boundary ($X, R, M$ points).
All compounds exhibit $M$-point instabilities. 
Excluding \ce{CsSnF3}, all compounds also exhibit $R$-point instabilities, and five of the eight systems exhibit $X$-point instabilities, \textit{viz.} \ce{CsSnF3}, \ce{CsSnI3}, \ce{CsPbCl3}, \ce{CsPbBr3}, and \ce{CsPbI3}. 
These zone boundary distortions are by definition antiferroelectric in nature, i.e. opposing polarisation induced in neighbouring unit cells cancels and no spontaneous polarisation is formed. 
In addition, all compounds excluding \ce{CsSnBr3} (Fig.\ref{phonon_band}(c) exhibit $\Gamma$-point instabilities, which is a ferroelectric instability that will not be considered further here. 

The presence of vibrational instabilities across all compositions is consistent with the scarcity of experimentally-observed cubic phases at low temperature (Table \ref{t:phase_table}) as anharmonic processes at high temperature are required for dynamic stabilization of the phase. 
In keeping with this, it is worth noting that from the present calculations cubic \ce{CsSnBr3}, which adopts a cubic structure close to room temperature, displays the smallest number of phonon instabilities amongst all the considered compositions. However, the number of imaginary modes is not necessarily related to the energetic barriers associated with the phase transition, which is the subject of the following section.

\begin{figure}
	\includegraphics{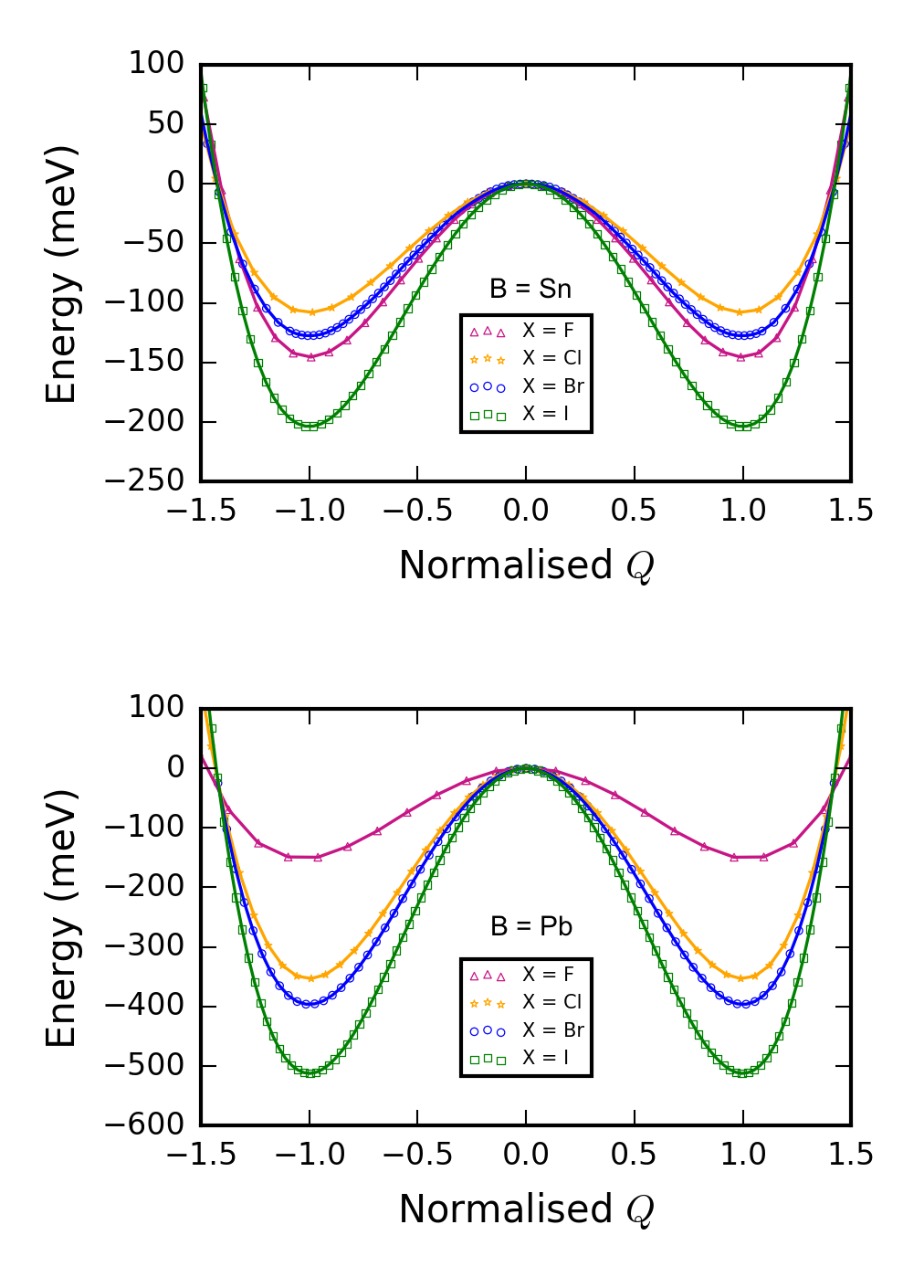}
	\caption{
    Double-well potential energy surface associated with the soft $M$-point modes in \ce{CsSnX3} (top), and \ce{CsPbX3} (bottom) from frozen-phonon calculations.
    The energy surface is calculated from DFT (2$\times$2$\times$2 supercell) and $Q$ is the distortion amplitude. 
    The energy zero refers to the undistorted structure.
	The high-symmetry cubic phase lies at $Q = 0$, and the distortion amplitude has been normalized so that the energetic minima lie at $Q = \pm 1$.
    The legend indicates the halide \ce{X} in each composition.
    }
	\label{double_well}
\end{figure}

\textit{Anharmonic potential energy surface}:
By distorting the crystal structure along a phonon eigenmode, the change in potential energy as a function of distortion amplitude (\textit{Q}) can be obtained.
For a harmonic phonon mode in an equilibrium structure, the change in energy with mode amplitude should be parabolic with the minimum at $Q = 0$. 
Double-well potential-energy surfaces are observed in each of the cubic perovskites studied here, which is consistent with anharmonic behaviour that can be described within Landau's theory of phase transitions.\cite{Dove1997}
The potential energy surface is well fitted by a function of the form:
\begin{equation}
E(Q) = aQ^2 + bQ^4 + O(Q^6)
\end{equation}
where $a$ and $b$ are fitted coefficients, and the former corresponds to the square of the harmonic phonon frequency.
For imaginary modes, representing structural instabilities, $a$ will be negative as the energy surface forms a double well with $Q = 0$ as a saddle point.

Mapping and fitting the anharmonic potential energy surfaces provides access to a number of quantities, including the depth of the well ($\Delta E$), the normal-mode coordinate of the local minima ($\Delta Q$), and the curvature of the potential energy about $Q = 0$.
The well depth $\Delta E$ determines the energy difference between the cubic and lower-symmetry structures represented by the distortion, and further dictates the transition rate between equivalent symmetry-broken distorted structures.
$\Delta Q$ determines the degree of distortion that minimises the potential energy. 
There are some caveats to this approach, as follows. 
The phonon eigenvector represents atomic motion by a three-component vector of orthogonal displacements, 
which cannot fully describe rotational motion.
For soft modes involving octahedral tilting, at large $Q$ the mode eigenvector may no longer accurately reflect the atomic displacements.
In practice, this means that a structure at the local minimum along the mode potential may undergo further relaxation if allowed to optimise freely, which would produce a larger energy barrier.
Secondly, in the case of degeneracy, linear combinations of the eigenvectors are valid solutions to the harmonic problem, and so the mode eigenvectors are not uniquely defined, and the ``true'' energy minimum may lie at a combination of the two.
In high-symmetry structures, however, the displacement pathways may be fixed by crystal symmetry.

Since all eight halide-perovskite compositions studied here exhibit singly-degenerate soft modes at the $M$-point, we take this as a representative instability and investigate these imaginary modes further.
The soft-mode potential wells of the eight compounds are plotted in Fig.\ref{double_well}.
To enable a direct comparison, the energy is calculated as a function of a normalised mode amplitude $Q$, such that the energy minima for each compound lie at $Q = \pm1$.
Across the \ce{CsPbX3} series, the well depth increases (i.e. the distorted structure lowers in energy relative to the cubic phase) systematically from \ce{F} to \ce{I}, with the \ce{Cl} and \ce{Br} perovskites having similar depths.
The same trends are evident in the \ce{CsSnX3} family from \ce{Cl} to \ce{I}, while \ce{CsSnF3} marks a notable exception, with a well depth comparable to those of the \ce{Cl} and \ce{Br} perovskites.
In addition, \ce{CsPbX3} appear to have deeper minima than \ce{CsSnX3} in general (excluding the anomaly \ce{CsSnF3}).

\begin{figure}
	\includegraphics{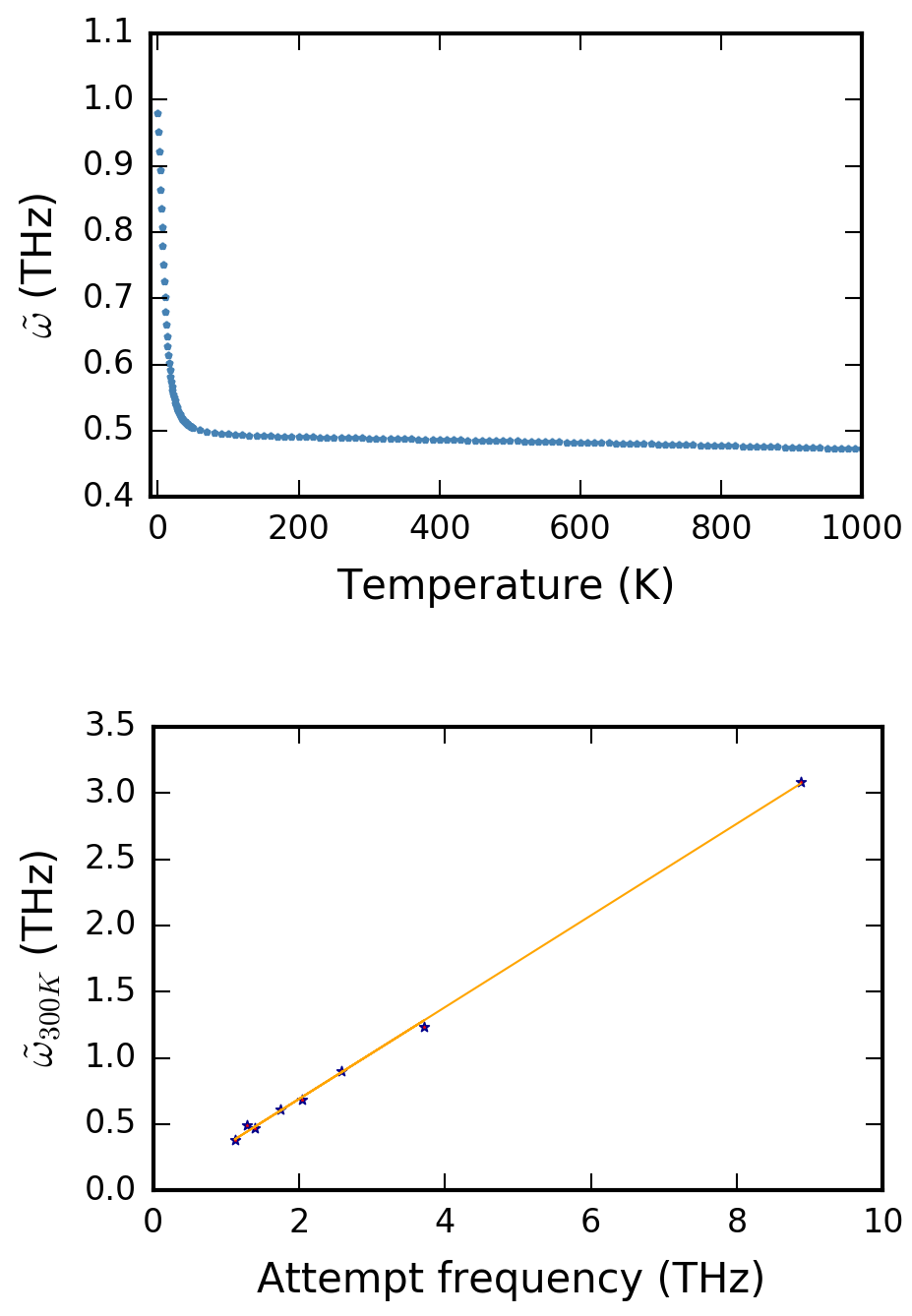}
	\caption{ (top) The renormalised phonon frequency ($\omega$) for the \textit{M} mode instability in \ce{CsPbI3} from a solution of the 1D Schr\"{o}dinger equation with the well depth of 512.3 meV.
	(bottom) Comparison of the renormalized effective phonon frequency ($\omega_{300K}$)  and the classical well curvature (attempt frequency) for  the \textit{M} mode instability of all inorganic halide perovskites considered at T = 300 K.  }
	\label{t_omega}
\end{figure}

\textit{Static or dynamic disorder}:
The rate of hopping between the symmetry equivalent local minima in the potential energy surface can be estimated in several ways. 
Firstly, we consider a classical kinetic model to compute a hopping rate ($\Gamma$) for the structural transition:
\begin{equation}
\Gamma = \nu \exp(\frac{-\Delta E}{k_B T})
\end{equation}
where $\nu$ is the attempt frequency that is equivalent around the curvature of the double well minima, $k_B$ is the Boltzman constant and $T$ is temperature. 
Alternatively, we can solve a Schr\"{o}dinger equation for the double well potential and define an effective harmonic frequency that reproduces the partition function  of the anharmonic system. Here we follow the procedure of Skelton et al. that has previously been applied to SnSe.\cite{Skelton2016,Buckeridge2011} 
The temperature dependence of the renormalised harmonic frequency is shown for \ce{CsPbI3} in Fig.\ref{t_omega}(a). 
It is found that the frequencies calculated using the (athermal) classical and (finite temperature) 
quantum approaches are in good agreement subject to a scaling factor (shown in Fig.\ref{t_omega}(b)).
At T = 300 K, the renormalization factor is 0.34, and the quantum solution suggests a characteristic vibration of 0.5 -- 3 THz depending on the chemical composition. 
The associated hopping rate for each compound is summarised in Table \ref{t:tunneling}. 
We note that rates are based on a single anharmonic mode and neglect phonon-phonon interactions that could be considered using a 
higher level of theory, e.g. a self-consistent phonon procedure.\cite{Koehler1966,Koehler1968,Souvatzis2008} 

For deeper wells, the transition between equivalent symmetry-broken local minima becomes less probable at a given temperature, giving rise to a slower hopping rate. 
Diffraction samples long-range order, within the penetration depth of the coherent beam. We assume that only if the transition rate between structures is $< 1$ Hz would the material phase segregate into macroscopically ordered domains. Such slow transitions are predicted for \ce{CsPbCl3}, \ce{CsPbBr3} and \ce{CsPbI3} at 150 K. We thus conclude that this dynamic disorder would not be observed with X-ray diffraction at room temperature. The random (non-correlated) orientations give rise to the observed higher-symmetry spacegroup.\cite{Dove1997}
 
Optoelectronic processes such as light absorption (fs), carrier thermalisation (fs), carrier scattering (sub ps) and recombination (ns), are all relatively fast. An adiabatic approximation can therefore be made. From the perspective of electrons, the potential energy surface is stationary with fixed distortions randomly orientated throughout the bulk. Electronic and optical processes sample the local symmetry broken structure, and thus its influence should be included in quantitative models of transport and device operation. 
The classification between static and dynamic disorder depends on the timescale of the interactions. 
In the limit of very high temperatures (i.e. $k_B T >> \Delta E$), all the halide perovskites would revert to dynamic disorder, although for the systems with large $\Delta E$ this limit would be much higher than the typical operating temperatures of semiconductor devices.

\begin{table}[]
\centering
\caption{Calculated transition rate  ($\Gamma$) across the double well potential (values in Hz). 
As the temperature increases, the hopping rate increases, but the absolute value also depends on the attempt frequency ($\nu$). 
Contributions from quantum mechanical tunnelling are not considered.
Note that the timescale of typical diffraction experiments is seconds, whilst electronics processes (carrier transport and recombination) can occur on timescales of $10^{-15} - 10^{-9}$ seconds.
}
\label{t:tunneling}
\begin{tabular}{cccccc}
\hline 
& $\Delta E$ (meV) & $\nu$ (THz) & $\Gamma$ at 150 K (Hz) & $\Gamma$ at 298 K (Hz) & $\Gamma$ at 500 K (Hz) \\ \hline
\ce{CsSnF3}    & 144.7    & 8.87   & $1.19\times10^{8}$   & $3.18\times10^{10}$   & $3.09\times10^{11}$ \\ \hline
\ce{CsSnCl3}   & 108.1    & 2.04   & $4.69\times10^{8}$   & $3.05\times10^{10}$   & $1.66\times10^{11}$ \\ \hline
\ce{CsSnBr3}   & 127.7    & 1.28   & $6.45\times10^{7}$   & $8.93\times10^{9}$    & $6.64\times10^{10}$ \\ \hline
\ce{CsSnI3}    & 203.7    & 1.12   & $1.55\times10^{5}$   & $4.04\times10^{8}$    & $9.90\times10^{9}$ \\ \hline
\ce{CsPbF3}    & 151.8    & 3.71   & $2.88\times10^{7}$   & $1.01\times10^{10}$   & $1.10\times10^{11}$ \\ \hline
\ce{CsPbCl3}   & 353.4    & 2.58   & $3.27$               & $2.76\times10^{6}$    & $7.10\times10^{8}$ \\ \hline
\ce{CsPbBr3}   & 396.5    & 1.74   & $7.81\times10^{-2}$  & $3.48\times10^{5}$    & $1.76\times10^{8}$ \\ \hline
\ce{CsPbI3}    & 512.3    & 1.40   & $7.91\times10^{-6}$  & $3.08\times10^{3}$    & $9.62\times10^{6}$ \\ \hline
\end{tabular}
\end{table}

Interestingly, we found that the local minima for \ce{CsSnF3} occurred at a relatively small absolute values of $Q$, suggesting that this perovskite might undergo a different type of $M$-point distortion to the other compounds.
On examining the eigenvectors, we verified that this is indeed the case: the $M$-point soft mode in \ce{CsSnF3} is an $M_2^-$ distortion, while those in other compositions correspond to $M_3^+$ tilts.
The $M_2^-$ mode is a second-order Jahn-Teller distortion where \ce{B-X} bonds shorten and lengthen, whereas the $M_3^+$ mode represents rigid in-phase octahedral tilting.
This can be explained by orbital mixing between the \ce{Sn} \textit{5s} and \ce{F} \textit{2p} orbitals, which produces the asymmetric electron density required to support a Jahn-Teller distortion.\cite{Walsh2011}

\begin{figure}
	\includegraphics{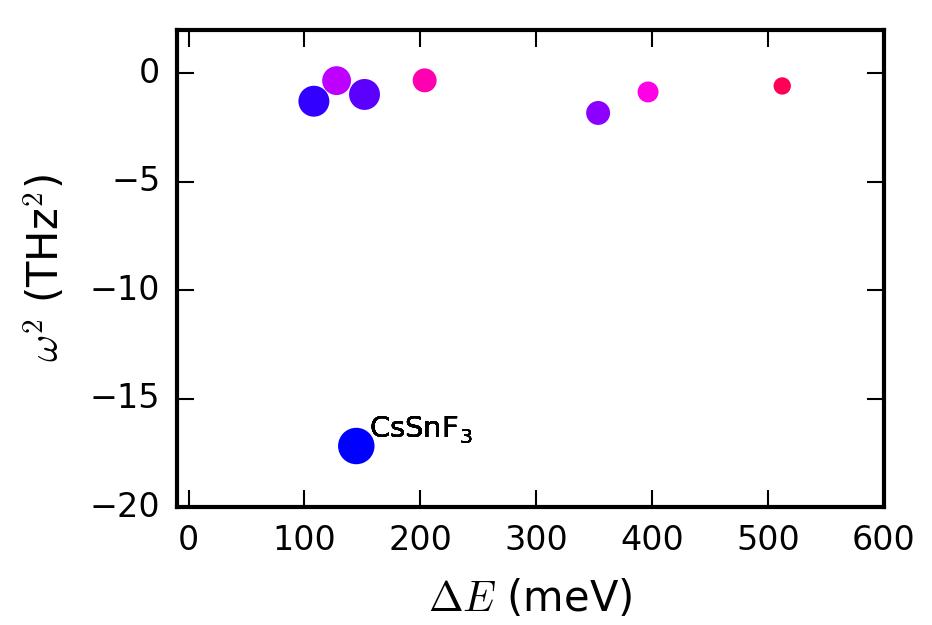}
	\caption{
    Squared imaginary harmonic phonon frequencies, $\omega^2$ (THz$^2$), for the $M$-point soft modes against the well depth (meV). 
    The size of the markers is proportional to the structural tolerance factors $\alpha$, which range from 0.93 (\ce{CsSnF3}) to 0.85 (\ce{CsPbI3}). The colours corresponds to the overall mass of the unit cell, with the heaviest being the ``warmest" (\ce{CsPbI3}: red) and the lightest the ``coolest'' (\ce{CsSnF3}: blue). 
    }
	\label{frequency_depth}
\end{figure}

\textit{Correlation between the harmonic frequency and $\Delta E$:}
The imaginary harmonic phonon frequency (at the saddle point) has been assumed to be indicative of the energetic driving force for distortions in perovskites.\cite{Benedek2013,Xiang2014}
To assess the correlation between the frequency and the depth of the minima, we plotted the well depths of the $M$-point soft modes obtained by the potential-energy mapping against the squared harmonic frequency $\omega^2$ (Fig. \ref{frequency_depth}).
From this analysis, we see that there is little correlation between the two.
Excluding the outlier \ce{CsSnF3}, the well depths span a 400 meV range, with a spread in $\omega^2$ of 1.5 THz$^2$.
The data are scattered across the energy range, with no clearly-evident patterns of frequency distribution, indicating that the imaginary harmonic frequencies may be a poor proxy for the well depths.
This can be understood from the fact that the harmonic frequency reflects the curvature of the potential-energy surface at the average structure ($Q = 0$), which does not contain sufficient information to extrapolate to the anharmonic region of the soft-mode potential.

Analysis of Figure \ref{frequency_depth} does, however, reveal a correlation between the well depth and the tolerance factor, namely that compounds with larger tolerance factors tend to produce shallower minima (i.e. the distorted structures are closer in energy to the cubic average structure).
This supports the established simple relationship between chemical composition and structural instability: the closer the tolerance factor to unity, the more ``cubic'' a structure is expected to be.
Our data shows that lower-symmetry configurations are indeed more energetically favorable for compositions with smaller tolerance factors.

\begin{figure}
	\includegraphics{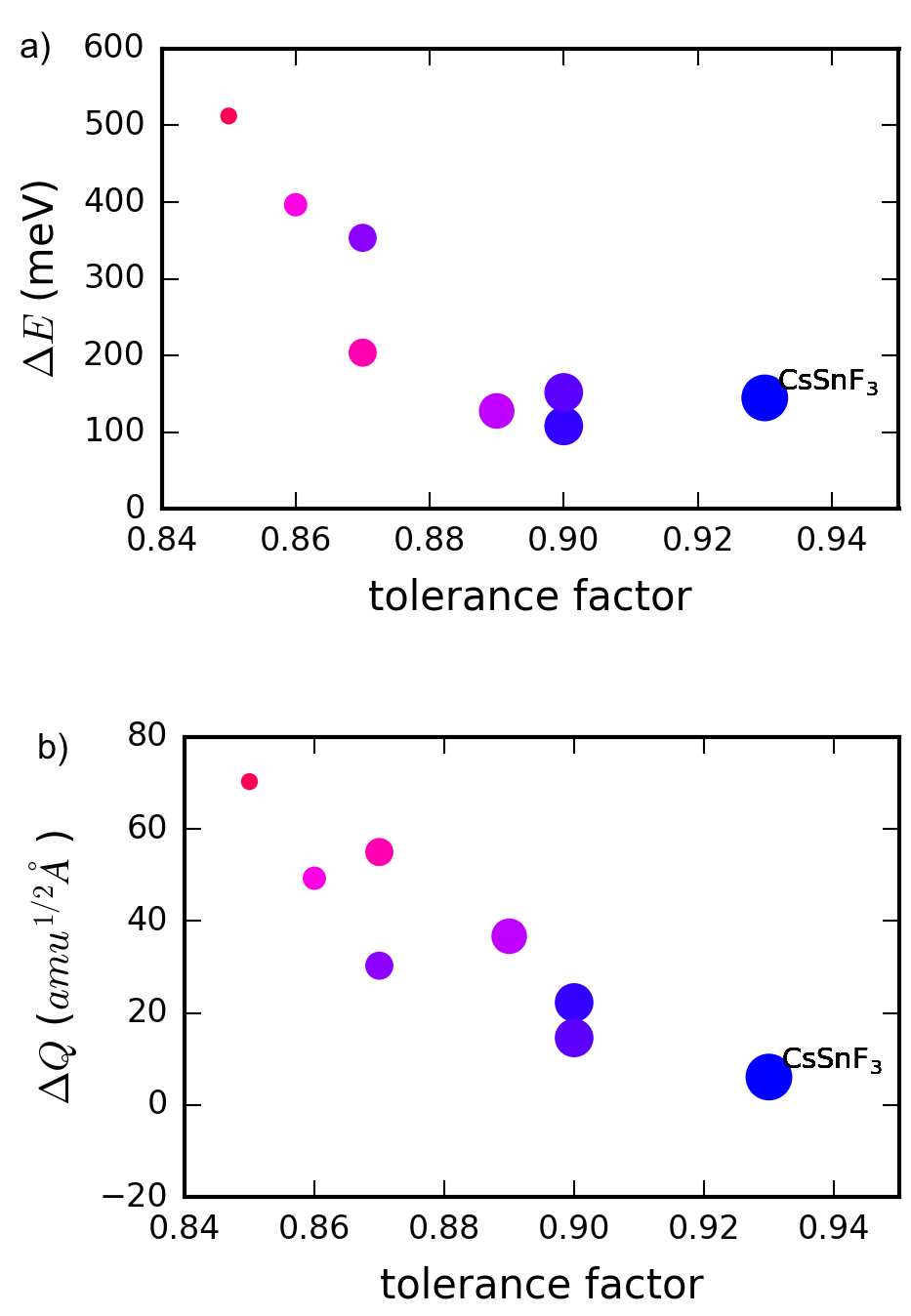}
	\caption{
    Well depths and distortion amplitudes along the $M$-point soft modes as a function of tolerance factor.
    The ionic radii used for each of the elements are: \ce{Cs}: 1.88 \AA; \ce{F}: 1.33 \AA; \ce{Cl}: 1.81 \AA; \ce{Br}: 1.96 \AA; \ce{I}: 2.20 \AA; \ce{Pb}: 1.19 \AA;\cite{Shannon1976} \ce{Sn}: 1.10 \AA.\cite{Li2008} The size and the color of the circles represent the tolerance factor and relative formula mass, as in Fig.\ref{frequency_depth}.}
	\label{Q_depth_t}
\end{figure}

For the \ce{CsPbX3} series, when \ce{X} increases from \ce{F} to \ce{I} the tolerance factor decreases from 0.90 to 0.85, and the temperature of the cubic phase transition increases from 187 K to 328 K, 413 K and 634 K.\cite{Berastegui2001,Harada1976,Moller1958,Trots2008}
This indicates that for compounds with smaller tolerance factors more thermal energy is required to lift the symmetry to cubic phase, which is in agreement with our calculations.
Structurally, when $\alpha <$ 1, the \ce{A-X} bonding is undercoordinated, and octahedral tilting is required to optimize the chemical bonding environments
If we explicitly plot the well depth as a function of tolerance factor (shown in Fig. \ref{Q_depth_t}a), this trend becomes apparent.
There is also a correlation between tolerance factor and the distortion amplitude that minimises the energy ($\Delta Q$), shown in Fig. \ref{Q_depth_t}b.
Compositions with small $\alpha$ require a larger distortion to the local minimum in order to optimize the cation-bonding environment due to the under coordination.

\begin{figure}
	\includegraphics{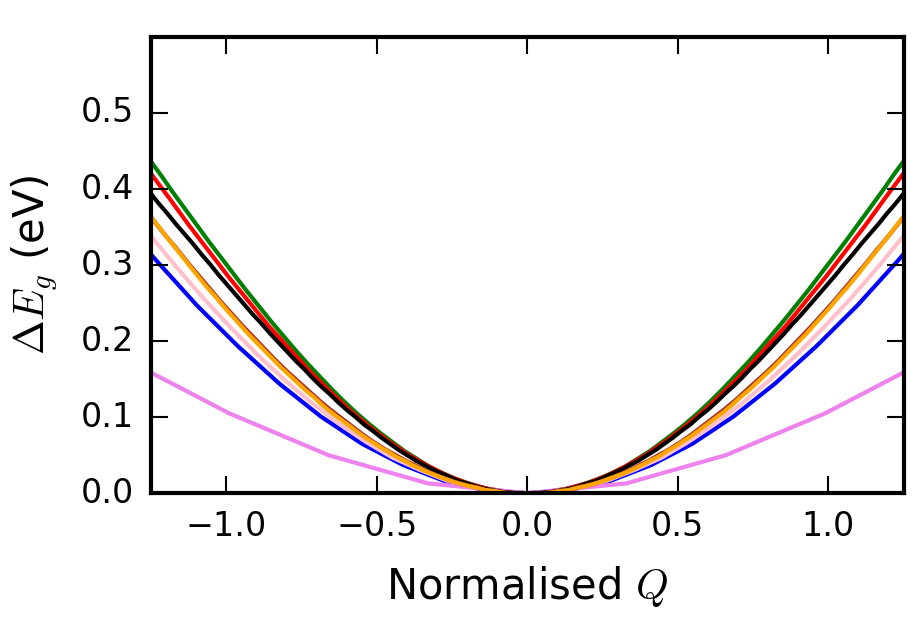}
	\caption{
    Change in band gap $\Delta E_g$ (eV) relative to the cubic structure for each compound as a function of normalised soft mode amplitude $Q$.
    The average cubic structure lies at $Q = 0$, while the lowest-energy distorted structure lies at $Q = 1$. The pink line, showing least change in $E_g$, corresponds to \ce{CsPbF3}, and green line, showing most change, corresponds to \ce{CsPbCl3}. 
    }
	\label{band_gap}
\end{figure}

\textit{Electronic structure effects:}
We further assess the effect of the $M$-point tilting distortions on the electronic structure by calculating the change in the band gap along the normal-mode coordinate $Q$.
The band gap for all eight compounds increase to different extents when distorting along the soft modes (Fig. \ref{band_gap}).
This indicates that distortions from average cubic symmetry will lead to a band gap increase in these perovskites. 
Although the semi-local exchange-correlation functional used to estimate the band gap underestimates the absolute value of the band gap , the relative shifts should be reliable. 
Due to the fact that the upper valence band consists of strong \ce{Sn}/\ce{Pb} \textit{s} and \ce{X} \textit{p} anti-bonding character, upon tilting the overlap of the orbitals decreases and results in a lower-energy valence-band maximum, and thus an increase in the gap.\cite{Woodward1997a,Huang2013c,Meloni2016}
The ultimate effect of such distortions on the band gap will also depend on the type of disorder.
Static disorder would lead to a more pronounced widening the band gap, whereas dynamic disorder would produce a less-pronounced time-averaged effect.
A more quantitative description will require the development and application of more sophisticated theories for anharmonic electron-phonon coupling. 

In summary, we have performed a comprehensive investigation of the phonon stabilities in the cubic \ce{CsSnX3} and \ce{CsPbX3} halide perovskites (X = F, Cl, Br, I).
Our results show that all eight compounds exhibit phonon soft-mode instabilities in the cubic phase.
Examining the potential energy surface along a representative soft-mode structural distortion reveals a correlation between the chemical composition and structural tolerance factor and the energetic barrier to accessing the high-symmetry structure.
We also found that the nature of the distortions differ depending on the chemistry, with \ce{CsSnF3} showing a fundamentally different $M$-point distortion to the other seven halide perovskites.
The hopping rate of the structural transition between two local minima is calculated which relates to the timescale of diffraction measurement.  
Finally, we have also studied the effect of the distortion from cubic symmetry on the electronic structure, and find that, in all cases, distortion along the $M$-point soft modes leads to a widening of the band gap.
There are implications on future electronic-structure studies and assessment of the role of local symmetry breaking and  electrostatic (band gap) fluctuations on the performance of perovskite optoelectronic devices. 

\section{Acknowledgement}
The research has been funded by the European Research Council (project 27757) and 
the EPSRC (grant nos. EP/K016288/1 and EP/K004956/1). 
Via our membership of the UK's HEC Materials Chemistry Consortium, which is funded by EPSRC (EP/L000202), this work used the ARCHER UK National Supercomputing Service (http://www.archer.ac.uk).

\bibliography{library}

\end{document}